\documentclass[a4paper,12pt]{article}

\usepackage[utf8]{inputenc}
\usepackage{amsfonts,amsmath,amssymb}
\usepackage{mathtools}
\usepackage{xcolor}
\usepackage{geometry}

\usepackage{authblk}
\usepackage{cite}
\usepackage[normalem]{ulem}
\usepackage{hyperref}

\usepackage{graphicx}
\usepackage{booktabs}

\usepackage{dutchcal}
\usepackage{extarrows}

\title{Lyapunov function for non-equilibrium\\transport processes}

\author[a]{Chuan-Jin Su}
\author[a]{Yu-Chao Hua\thanks{Corresponding author. \textit{E-mail:} \href{mailto:huayuchao19@163.com}{huayuchao19@163.com}}}
\author[a]{Zeng-Yuan Guo}

\affil[a]{Key Laboratory for Thermal Science and Power Engineering of Ministry of Education, Department of Engineering Mechanics, Tsinghua University, Beijing 100084, China.}

\date{}

\begin{document}

\maketitle

\begin{abstract}
Irreversibility is a critical property of non-equilibrium transport processes. An opinion has long been insisted that the entropy production rate is a Lyapunov function for all kinds of processes, that is, the principle of minimum entropy production. However, such principle is based on some strong assumptions that are rarely valid in practice. Here, the common features of parabolic-like transport processes are discussed. A theorem is then put forward that the dot products of fluxes and corresponding forces serve as Lyapunov function for parabolic-like transport processes. Such fluxes and forces are defined by their actual constitutive relations (e.g., the Fourier's law, the Fick's law, etc.). Then, some typical transport processes are analyzed. Particularly for heat conduction, both the theoretical and numerical analyses demonstrate that its Lyapunov function is the entransy dissipation rather that the entropy production, when the Fourier's law is valid. The present work could be helpful for further understanding on the irreversibility and the mathematical interpretation of non-equilibrium processes.
\end{abstract}

\section{Introduction}
Investigation of transport processes is one crucial topic in non-equilibrium thermodynamics \cite{demirel_nonequilibrium_2019}. In practice, linear constitutive relations are generally adopted to handle the non-equilibrium transport problems \cite{bird2006transport}, which lead to parabolic-like governing equations. Thus, the transport processes characterized by parabolic-like partial differential equations (PDEs) can be called parabolic-like transport processes.

The understanding of irreversibility which is a fundamental property of non-equilibrium transport processes, plays an important role in the study of transport processes. For the parabolic-like processes, irreversibility is reflected by governing equations where the partial time derivative terms invalidate the time reversal symmetry. In order to propose a quantity that measures irreversibility, the concept of Lyapunov function in modern stability theory is introduced into non-equilibrium thermodynamics, as irreversibility is the expression of the attraction of stability, quoted by Prigogine \cite{prigogine1980being}. For any system, a scalar function whose time derivative has the opposite is called a Lyapunov function, and the system is stable in the sense of Lyapunov \cite{la2012stability}. Classical stability theory formulated by Gibbs for equilibrium states is based on the variation of entropy, where the second term is negative, i.e. $\frac{1}{2}\delta^2 S<0$ \cite{kondepudi2014modern}, however, the similar methodology cannot be applied for the non-equilibrium stationary states because of the limitations of such variational formulations out of equilibrium \cite{lebon2008understanding}.

Since 1940s, I. Prigogine has proposed and further developed the principle of minimum entropy production (MinEP) \cite{prigogine1945etude,prigogine1965introduction} as the key criterion of stability for non-equilibrium transport processes, characterizing a system at stationary state by producing entropy at the lowest rate compatible with external constraints. This theorem holds only if the following hypotheses \cite{glansdorff1954} are satisfied:
\begin{enumerate}
	\item Linear relations between generalized forces and fluxes in irreversible phenomenon.
	\item Validity of Onsager reciprocal relations.\label{hypo:onsager}
	\item Phenomenological coefficients are considered constant.\label{hypo:pheno_coeff_constant}
\end{enumerate}
When applying to a perturbed system that satisfies them all, we shall see that irreversible processes inside always lower the value of entropy production rate (EPR), and the system returns to the state at which its entropy production is lowest \cite{prigogine1980being}, if its boundary conditions are invariant of time.

However, questions arise in most practical cases. Firstly, the assumption that constitutive relations between thermodynamic fluxes and forces are entropic (i.e., their dot products should be the entropy production, suggested by previously listed hypotheses) is hardly corresponding to practical situations. As an example, for the heat conduction process, a strong assumption is required that the thermal conductivity must be inversely proportional to the square of the temperature \cite{zullo2016,martyushev2007}, i.e. $k=l_{qq}/T^2\propto\ 1/T^2$. However, there are few materials of which properties meet such assumption \cite{jaynes1980}. On the other hand, such assumption is equivalent to the validation of the principle of MinEP, which means that a circular logic is misused, and no additional information is provided by such minimum principle.

In this article, we put forward a theorem stating that for given constitutive relations, the dot products of thermodynamic fluxes and forces serve as Lyapunov functions for parabolic-like transport processes. Lyapunov functions for some typical transport processes are studied in section \ref{sec:typical}. For the heat conduction process under the Fourier's law, a Lyapunov function provided by the theorem is the entransy dissipation rate which is a measure of the heat conduction process' irreversibility \cite{guo2007entransy}, supporting the conclusion of Hua et al. \cite{hua2017,hua2018entropy}. Numerical calculations are also provided to support the theorem in section \ref{subsec:counterexamples}.

Applying the theorem, the principle of MinEP and its problematics are discussed in \ref{sec:misunderstanding}. In addition, the attempts to modify the principle of MinEP are reviewed, including Prigogine's approximation that the coefficient $l_{qq}$ should be constant in a small temperature range \cite{kondepudi2014modern}, and the non-total differential \cite{glansdorff1954} which stays non-positive and is claimed to be a general property of entropy production.

\section{Lyapunov function for parabolic-like transport processes}
\label{sec:parabolic_model}

As mentioned above, for a system, a scalar function whose time derivative has the opposite sign is a Lyapunov function \cite{la2012stability}. We will demonstrate that the dot products of fluxes and forces serve as Lyapunov functions for parabolic-like transport processes.

The local equilibrium assumption of parabolic-like transport processes take the form of the following balance equation \cite{lebon2008understanding},
\begin{equation}
	\rho \frac{\partial a}{\partial t} = -\nabla \cdot \boldsymbol{J}^a + \sigma^a,
	\label{eq:local_assumption}
\end{equation}
where $a$ is an extensive state variable under consideration, $\boldsymbol{J}^a$ is the related flux term, $\sigma^a$ is the corresponding source term; and $a$ is conserved when $\sigma^a$ is zero, which is the case that we consider in this paper.

\subsection{Linear transport processes}

For linear transport processes, constitutive relations between driving forces and fluxes are expressed,
\begin{equation}
	\boldsymbol{J}^a = K \boldsymbol{X^a} = K \nabla \Gamma^a,
	\label{eq:constitutive_relation}
\end{equation}
where $K$ is a constant, $\Gamma^a$ is the corresponding intensive state variable of $a$.

We now define $G$ as the integral of the dot product $g$ of a flux $\boldsymbol{J}^a$ and its driving force $\boldsymbol{X}^a$ within a system $V$,
\begin{equation}
	G^a = \int_V g^a \mathrm{d}V = \int_V \boldsymbol{X}^a \cdot \boldsymbol{J}^a\mathrm{d}V,
	\label{eq:G_product}
\end{equation}
whose sign is dependent on the constant $K$. For example, $G^e = \int \nabla (-T) \cdot \boldsymbol{q} =\int\boldsymbol{q}\cdot\boldsymbol{q}/k\geq 0$ for the heat conduction process, where $e$ is the specific thermal energy, $T$ is the temperature, $\boldsymbol{q}$ is the heat flux, and $k$ is a constant called the thermal conductivity, given the Fourier's law,
\begin{equation}
	\boldsymbol{q}=k \nabla (-T).
	\label{eq:fourier_law}
\end{equation}

If the chosen state quantity $a$ is conserved, the time derivative of $G^a$ can be immediately expressed using constitutive relation in eq. \eqref{eq:constitutive_relation},
\begin{equation}
	\begin{aligned}
		\frac{\mathrm{d}G^a}{\mathrm{d}t} &= \frac{\mathrm{d}}{\mathrm{d}t} \int_V \boldsymbol{X}^a \cdot \boldsymbol{J}^a\mathrm{d}V\\
		&= \frac{\mathrm{d}}{\mathrm{d}t} \int_V \nabla \Gamma^a \cdot K \nabla \Gamma^a \mathrm{d}V\\
		&= 2 \int_V K \nabla \Gamma^a \cdot \nabla \left(\frac{\partial \Gamma^a}{\partial t}\right)\mathrm{d}V\\
		&\xlongequal{\text{Gauss}} 2\int_A K \frac{\partial \Gamma^a}{\partial t}\frac{\partial \Gamma^a}{\partial \boldsymbol{n}} \mathrm{d}A - 2\int_V K\frac{\partial \Gamma^a}{\partial t} \nabla^2 \Gamma^a \mathrm{d}V,
	\end{aligned}
\end{equation}
where the first integral vanishes as the boundary conditions are fixed, i.e. the quantity $\Gamma^a$ is time independent on boundaries. Inserting the local equilibrium assumption in eq. \eqref{eq:local_assumption} without the source term, i.e. $\sigma^a=0$, it becomes,
\begin{equation}
	\frac{\mathrm{d}G^a}{\mathrm{d}t} = 2\int_V \rho \frac{\partial \Gamma^a}{\partial t}\frac{\partial a}{\partial t}\mathrm{d}V = 2\int_V \rho \frac{\partial \Gamma^a}{\partial a} \left(\frac{\partial a}{\partial t}\right)^2 \mathrm{d}V.
	\label{eq:time_derivative_G}
\end{equation}

For parabolic-like transport processes (e.g., the heat conduction process), the monotonic functional relation between the state quantities $a$ and $\Gamma^a$ is given as $\Gamma^a = \Gamma^a(a)$, which assures that the sign of $\frac{\partial \Gamma^a}{\partial a}$ is opposite to that of the constant $K$ ($\frac{\partial \Gamma^a}{\partial a} = -\frac{\partial T}{\partial u} = -\frac{1}{c_v} < 0$ for the heat conduction process, where $c_v$ is the specific heat capacity), governing equations can be expressed in the parabolic-like form,
\begin{equation}
	\rho\frac{\partial a}{\partial t} = -K\nabla\cdot\left(\frac{\partial \Gamma^a}{\partial a}\nabla a\right).
	\label{eq:parabolic_like_governing}
\end{equation}
We assert that for parabolic-like transport processes, the sign of such quantity $G^a$ is opposite to that of its time derivative $\frac{\mathrm{d}G^a}{\mathrm{d}t}$. Therefore, the quantities defined as $G$ serve as Lyapunov functions in such irreversible processes.

\subsection{Non-linear transport process}

When the constitutive relations between fluxes and corresponding forces are non-linear,
\begin{equation}
	\boldsymbol{J}^a = K(\Gamma^a)\nabla \Gamma^a,
\end{equation}
where the conductivity term $K$ is $\Gamma^a$-dependent, such relations can be rewritten in a linear form by applying the methodology in literature \cite{hua2017,chen1982heat} which was proposed to convert the problem of the temperature dependence of the thermal conductivity. By defining a generalized quantity
\begin{equation}
	F(\Gamma^a) = \int_{\Gamma^a_0}^{\Gamma^a}\frac{K(\Gamma^{a*})}{K_0}\mathrm{d}\Gamma^{a*},
\end{equation}
where $\Gamma^a_0$ is the reference point, and $K_0\equiv K(\Gamma^a_0)$. The non-linear relation is converted into a linear problem and can be rewritten as
\begin{equation}
	\boldsymbol{J}^a = K_0 \nabla F,
\end{equation}
as $\nabla F = \frac{K(\Gamma^a)}{K_0}\nabla \Gamma^a$. A Lyapunov function can be given by the previous theorem due to the linear constitutive relation between the flux $\boldsymbol{J}^a$ and the generalized driving force $\boldsymbol{X}^a_\text{ge} = \nabla F$, that is
\begin{equation}
	G_\text{ge}^a = \int_V \boldsymbol{X}^a_\text{ge} \cdot\boldsymbol{J}^a \mathrm{d}V.
	\label{eq:gge}
\end{equation}

Therefore, in the case of $G_{\text{ge}}^a$ (eq. \eqref{eq:gge}) serving as the Lyapunov function, the time evolution equation is the parabolic-like transport model in eq. \eqref{eq:parabolic_like_governing}, and the approach is to the fix point of the stationary states observed in practice. It should be noted that, in the framework of GENERIC, when considering the reducing rate relation from the arbitrary non-equilibrium level $\mathcal{L}$ to the stationary level $L$, such linearized relation can also be expressed by the GENERIC equation \cite{grmela2018genericguide,grmela2019entropy,grmela2021multiscale}, where the dissipation potential is $\Psi = \frac{1}{2}K_0 (\nabla \mathcal{a}^{*})^2$, and the linearized conjugate of the quantity $a$ is defined as $\mathcal{a}^{*}=F$. A Lyapunov function can be obtained from the rate thermodynamic potential \cite{grmela2021multiscale} and is in the same form as $G_{\text{ge}}^a$ in this case.

\section{Lyapunov functions for some typical transport processes}
\label{sec:typical}
\subsection{The heat conduction process}

Given the linear constitutive relation, i.e. the Fourier's law for the heat conduction process,
\begin{equation}
	\boldsymbol{q}=k \nabla (-T),
	\tag{\ref{eq:fourier_law} revisited}
\end{equation}
where $\boldsymbol{q}$ is the heat flux, and $\boldsymbol{X}^e = \nabla(-T)$ is the corresponding driving force. Applying our theorem of Lyapunov function for parabolic-like transport processes, a Lyapunov function is expressed as
\begin{equation}
	G^e = \int_V \boldsymbol{X}^e \cdot \boldsymbol{q} \mathrm{d}V = \int_V k \nabla T\cdot \nabla T\mathrm{d}V.
	\label{eq:lya_heat_cond}
\end{equation}

Guo et al. \cite{guo2007entransy} introduced entransy to characterize the ability of heat transfer. The local entransy dissipation rate of heat conduction process is the dot product of heat flux and the negative of temperature gradient,
\begin{equation}
	g = -\boldsymbol{q}\cdot\nabla T.
\end{equation}
The entransy dissipation in solid is a measure of the heat conduction process' irreversibility \cite{guo2007entransy}, and a minimum principle can be constructed by the entransy dissipation rate \cite{hua2017,hua2018entropy}. It should be noted that, the Lyapunov function $G^e$ defined by the theorem is exactly the entransy dissipation rate,
\begin{equation}
	G^e = - \int_V g \mathrm{d}V \geq 0,\qquad \frac{\mathrm{d}G^e}{\mathrm{d}t} \leq 0.
\end{equation}
where the equality holds if and only if stationary state is reached. Therefore, Lyapunov conditions for the heat conduction process are given by the entransy dissipation rate $G^e>0$ and its time derivative $\frac{\mathrm{d}G^e}{\mathrm{d}t}<0$.

Least generalized entransy dissipation principle is proposed by Hua and Guo \cite{hua2017}, to extend the scope of the entransy concept to the case of temperature dependent thermal conductivity. By defining a generalized temperature,
\begin{equation}
	F(T) = \int_{T_0}^T \frac{k(T')}{k_0(T_0)}\mathrm{d}T',
\end{equation}
where $T_0$ is the reference temperature. The generalized entransy dissipation rate is then defined as,
\begin{equation}
	G_\text{ge}^e = -\int_V\boldsymbol{q}\cdot\nabla F\mathrm{d}V = \int_V k_0 (\nabla F)^2\mathrm{d}V,
\end{equation}
which is the dot product of the heat flux and its generalized driving force $\boldsymbol{X}^e_\text{ge} = \nabla(-F)$. Indeed, $G_\text{ge}^e$ serves as a Lyapunov function, as pointed out by the non-linear case of the theorem.

\subsection{The mass diffusion process}

The Fick's law of diffusion describes the mass diffusion process, relating the diffusion flux to the gradient of the concentration,
\begin{equation}
	\boldsymbol{J}_\alpha = D_\alpha \nabla (-\phi_\alpha), \label{eq:fick_law}
\end{equation}
where $\boldsymbol{J}_{\alpha}$ is the diffusive flux of component $\alpha$, $D_{\alpha}$ is the diffusivity, $\phi$ is the concentration. A Lyapunov function for the molecular diffusion process is provided by the dot product of the flux and the corresponding driving force $\boldsymbol{X}^\phi = \nabla(-\phi_\alpha)$,
\begin{equation}
	G^\phi = \int_V \boldsymbol{X}^\phi \cdot \boldsymbol{J}^\phi \mathrm{d}V = \int_V D_\alpha \nabla \phi_\alpha \cdot \nabla \phi_\alpha\mathrm{d}V
\end{equation}
with the conditions $G^\phi \geq 0$ and $\frac{\mathrm{d}G^\phi}{\mathrm{d}t}\leq 0$.

\subsection{The electrical conduction process}

The electrical conduction process in electrical conductors is well described by the Ohm's law, whose vector form is expressed as,
\begin{equation}
	\boldsymbol{J}^q=\sigma \nabla (-V),
\end{equation}
where $\boldsymbol{J}^q$ is the current density, the constant $\sigma$ is the conductivity, and $V$ is the electric potential. The superscript $q$ represents the charge density.

A Lyapunov function is written as the dot product of the flux $\boldsymbol{J}^q$ and its driving force $\boldsymbol{X}^q=\nabla (-V)$,
\begin{equation}
	G^q = \int_V \boldsymbol{X}^q \cdot\boldsymbol{J}^q = \int_V \sigma \nabla V\cdot\nabla V\mathrm{d}V,
\end{equation}
which is the Joule heating power within the system $V$, i.e. the rate of electrical energy dissipation. The previous theorem derives the same result as the common sense that the Joule heating power $G^q$ decreases to the minimum possible value for given boundary conditions \cite{jaynes1980}.

\begin{table}[htbp]
	\centering
	\caption{Constitutive relations and Lyapunov functions of some typical transport processes}
	\begin{tabular*}{\textwidth}{@{}@{\extracolsep{\fill}}lcl@{}}
		\toprule		
		Transport process & Constitutive relation & Lyapunov function\\
		\midrule
		Thermal conduction & $\boldsymbol{q}=k\nabla(-T)$ & $G^e=\int k\left(\nabla T\right)^2$\\
		Mass diffusion & $\boldsymbol{J}_\alpha = D_\alpha \nabla (-\phi_\alpha)$ & $G^\phi =\int D_\alpha\left( \nabla \phi_\alpha\right)^2$\\
		Electrical conduction & $\boldsymbol{J}^q=\sigma \nabla (-V)$ & $G^q = \int \sigma \left(\nabla V\right)^2$\\
		\bottomrule
	\end{tabular*}
\end{table}

The defined Lyapunov functions $G^e, G^\phi$ and $G^q$ suggest the stability of these typical transport processes, and measure the irreversibility featured by the parabolic-like governing equations. Since the validation of the theorem is verified, it should be pointed out that the fluxes and their corresponding forces should be defined by actual constitutive relations as above. Such $G$-type Lyapunov function can be also used in the framework of rate thermodynamics \cite{grmela2021multiscale} to characterize the overall pattern of the preparation process from non-equilibrium state to the non-equilibrium stationary state.

In addition, it is worth mentioning that for the equilibrium states, both the function $G^a$ and the entropy $S$ express the approach to the equilibrium, as the final stage can be defined by both equilibrium state and non-equilibrium stationary state. And the Lyapunov function of type $G$ exist even in externally driven systems without local equilibrium for which the entropy $S$ does not exist \cite{grmela2021multiscale}.

\section{Misunderstandings on the entropy production and its minimum principle}
\label{sec:misunderstanding}
\subsection{Limitations of the principle of MinEP}

It has been mistakenly believed that the entropy production rate (EPR) is a Lyapunov function for any non-equilibrium transport processes, since the principle of minimum entropy production (MinEP) was proposed by I. Prigogine \cite{prigogine1945etude,prigogine1965introduction}, stating that a system at stationary state by producing entropy at the lowest rate compatible with external constraints, and the EPR plays the role of Lyapunov function to characterize the stability of such processes. It is crucial to point out that, the principle of MinEP is valid only when following entropic constitutive relations are given, which invalidate the applicability and the universality of the principle. For the heat conduction process and the mass diffusion process discussed above,
\begin{align}
	\boldsymbol{q} &= l_{qq} \nabla \left(\frac{1}{T}\right),\\
	\boldsymbol{J}_{\alpha} &= l_{\phi\phi} \nabla \left(-\frac{\mu_\alpha}{T}\right),
\end{align}
where $l_{qq}$ and $l_{\phi\phi}$ are phenomenological coefficients in such sense, $\mu_\alpha$ is the chemical potential of component $\alpha$. The driving forces are defined as
\begin{equation}
	\boldsymbol{X}^u_{\text{ep}} = \nabla \left(\frac{1}{T}\right),\qquad \boldsymbol{X}^{\phi}_{\text{ep}}=\nabla \left(-\frac{\mu_\alpha}{T}\right),
\end{equation}
which make the dot products of fluxes and corresponding forces to be the EPR. Lyapunov functions for such processes are given by our previous theorem,
\begin{align}
	P_u &= \int_V \boldsymbol{J}^u\cdot\boldsymbol{X}^u_{\text{ep}} \mathrm{d}V = \int_V \boldsymbol{q}\cdot\nabla\left(\frac{1}{T}\right)\mathrm{d}V,\label{eq:entropy_conduction}\\
	P_\phi &= \int_V \boldsymbol{J}^\phi \cdot\boldsymbol{X}^\phi_{\text{ep}} \mathrm{d}V = \int_V \boldsymbol{J}_{\alpha}\cdot \nabla \left(-\frac{\mu_\alpha}{T}\right)\mathrm{d}V,\label{eq:entropy_diffusion}
\end{align}
which confirm the conditional validation of the principle of MinEP (see also \cite{demirel_nonequilibrium_2019,zullo2016,degroot2013} for the heat conduction process).

\subsection{The principle's problematics}

However, there are two issues if the EPR is considered to be a Lyapunov function. Firstly, the hypothesis of constant phenomenological coefficients is hardly corresponding to practical situations. As an example, for the heat conduction process, a strong assumption is required that the thermal conductivity must be inversely proportional to the square of the temperature \cite{zullo2016,martyushev2007}, i.e., $k=\frac{l_{qq}}{T^2}\propto \frac{1}{T^2}$. However, there are few materials of which properties meet such assumption \cite{jaynes1980}. When discussing entropy production in electrical circuit elements, Kondepudi and Prigogine \cite{kondepudi2014modern} provided the entropic constitutive relation between the voltage and the current as following,
\begin{equation}
	I = L_R \frac{V_R}{T},
\end{equation}
where $L_R$ is a phenomenological coefficient. The same problem arises, that is the resistance $R=\frac{T}{L_R}$ should be proportional to the temperature, which is not the generally adopted case where the Ohm's law is valid.

It should be noted that the principle of MinEP will be violated if linear phenomenological relations (e.g., the Fourier's law) are valid. Take the heat conduction process as an example, the EPR is written as eq. \eqref{eq:entropy_conduction}, whose time derivative becomes,
\begin{equation}
	\begin{split}
		\frac{\mathrm{d}P_u}{\mathrm{d}t} &= \frac{\mathrm{d}}{\mathrm{d}t} \int_V \boldsymbol{q}\cdot\left(\nabla\frac{1}{T}\right)\mathrm{d}V\\
		& = \frac{\mathrm{d}}{\mathrm{d}t} \int_V k \nabla \ln T\cdot\nabla\ln T \mathrm{d}V\\
		&= 2 \left(\int_{A } k \frac{\partial \ln T}{\partial t} \frac{\partial \ln T}{\partial \boldsymbol{n}} \mathrm{d}A - \int_V k \frac{\partial \ln T}{\partial t} \nabla^2 \ln T \mathrm{d}V\right),
	\end{split}
	\label{eq:time_derivative_epr_kconst}
\end{equation}
applying the Gauss's divergence theorem. Given that the boundary conditions are fixed $\left.\frac{\partial T}{\partial t}\right|_A = 0$, insertion of the local equilibrium eq. \eqref{eq:local_assumption} into eq. \eqref{eq:time_derivative_epr_kconst} yields,
\begin{equation}
	\frac{\mathrm{d}P_u}{\mathrm{d}t} = - 2\int_V \frac{k^2}{\rho c} \frac{1}{T} \nabla^2 T \left(\frac{1}{T}\nabla^2 T - \frac{\nabla T\cdot\nabla T}{T^2}\right)\mathrm{d}V,
	\label{eq:time_derivative_epr_kconst_final}
\end{equation}
whose sign is uncertain. Hence, when the initial condition is chosen appropriately, the principle of MinEP can be surely violated (see a counterexample in section \ref{subsec:counterexamples}). If the same analysis is applied to the molecular diffusion process given that the Fick's law of diffusion is valid as eq. \eqref{eq:fick_law}, the same conclusion can be drawn as the time derivative of the EPR is expressed as,
\begin{equation}
	\frac{\mathrm{d}P_\phi}{\mathrm{d}t} = \int_V RD_\alpha^2 \frac{\nabla\phi_\alpha\cdot\nabla\phi_\alpha}{\phi_\alpha^2}\nabla^2 \phi_\alpha \mathrm{d}V.
\end{equation}
The principle of MinEP is violated when the inequality $\nabla^2 \phi_\alpha>0$ holds. As for the case of electrical circuit, the same problematic reappears: the principle of MinEP becomes invalid if the resistance is constant, that is, the Ohm's law rather than the entropic constitutive relation, are valid. A simple counterexample can be given by considering two parallel resistors at different temperatures \cite{jaynes1980}.

Secondly, a circular logic is misused. Given that the theorem holds, the validation of the principle of MinEP is equivalent to the assumption that constitutive relations between fluxes and forces should be entropic, which means that no additional information can be provided by such principle. Such logic of defining fluxes and their corresponding forces is completely different from that we applied above when studying specific transport processes.

\subsection{Prigogine's approximation on phenomenological coefficients}

Prigogine was aware of the issue of constant phenomenological coefficients, an approximation was made when discussing stationary states in the heat conduction process \cite{kondepudi2014modern},
\begin{equation}
	l_{qq} = kT^2 \approx kT_{avg}^2,
\end{equation}
where the coefficient $l_{qq}$ is treated approximately as constant as the average temperature $T_{avg}$ hardly changes too much. However, such approximation can be rejected as the time derivative of the EPR in eq. \eqref{eq:time_derivative_epr_kconst_final} can be positive no matter how small the temperature range is, as long as we put forward the corresponding initial temperature field.

Zullo \cite{zullo2016} points out that the assumption of constant average temperature $T_{avg}$ is not valid when the system is not isolated, which is the reason why this approximation is not applicable, and a further assumption should be made to neglect the derivative of $T_{avg}$. However, it should be clarified that even for an isolated system, the first integral in eq. \eqref{eq:time_derivative_epr_kconst} is still 0 as heat flux on the surface does not exist,
\begin{equation}
	\left. \frac{\partial T}{\partial \boldsymbol{n}}\right|_{A}=0.
\end{equation}
It is clear that for isolated systems, that is when the time derivative of average temperature $T_{avg}$ vanishes, the EPR is still inapplicable.

A point should be added concerning that approximation, that is in the expression of the EPR
\begin{equation}
	P = \int_V k\left(\frac{\nabla T}{T}\right)^2 \mathrm{d}V \approx \frac{1}{T_{avg}^2} \int_V k\left(\nabla T\right)^2 \mathrm{d}V,
\end{equation}
the average temperature $T_{avg}$ is a weighted average of the inverse of the gradient if we extract it outside the integral, which does dot have actual physical meaning. It would be pointless to make further assumptions on $T_{avg}$. It can be thus asserted that when the thermal conductivity is constant, the principle of MinEP simply does not hold.

\subsection{The non-total differential}

Glansdorff and Prigogine \cite{glansdorff1954} proposed a new methodology to answer the question of what the most general properties of entropy production are, independent of restrictive hypotheses. The non-total differential of the EPR is put forward to play a role of Lyapunov condition. Note $P$ the EPR, for fluxes and driving forces under entropic constitutive relations,
\begin{equation}
	P = \int_V \sum_j X_j J_j \mathrm{d}V,
	\label{eq:entropy_production_rate_sum}
\end{equation}
where $X_j$ and $J_j$ are thermodynamic forces and fluxes, enumerated by $j$. By decomposing time derivative of eq. \eqref{eq:entropy_production_rate_sum},
\begin{equation}
	\begin{split}
		\frac{\mathrm{d} P }{\mathrm{d} t} &= \int_V \sum_j \frac{\mathrm{d}X_j}{\mathrm{d}t}J_j \mathrm{d}V + \int_V \sum_j X_j \frac{\mathrm{d}J_j}{\mathrm{d}t} \mathrm{d}V\\
		&\equiv \frac{\mathrm{d}_X P}{\mathrm{d}t} + \frac{\mathrm{d}_J P}{\mathrm{d}t},
	\end{split}
\end{equation}
it can be obtained that the first component $\mathrm{d}_X P$ that always stays negative, or zero when reaching stationary state, even at least one of the hypotheses are invalid. Naive proofs of processes of the heat conduction, the chemical reaction, and the isothermal diffusion can be found in literature \cite{degroot2013,kondepudi2014modern,glansdorff1954}. In addition to the EPR itself $P\geq0$, another Lyapunov condition is given by its such non-total differential \cite{demirel_nonequilibrium_2019},
\begin{equation}
	\frac{\mathrm{d}_X P}{\mathrm{d}t} \equiv \int_V \sum_j \frac{\mathrm{d}X_j}{\mathrm{d}t}J_j \mathrm{d}V \leq 0.
	\label{eq:dxp}
\end{equation}

However, the so-called non-total differential is questionable. Firstly, the concept is ambiguous although a new `kinetic potential' was introduced by Kondepudi and Prigogine \cite{kondepudi2014modern} such that $\mathrm{d}W\equiv\mathrm{d}_X P$, as no explicit physical meaning has been interpreted so far. Secondly, the inequality \eqref{eq:dxp} holds not only for entropic constitutive relations. In fact, the properties of such defined non-total differential are universally valid for parabolic-like transport processes discussed previously, regardless of whether the constitutive relations are entropic. For any non-equilibrium process that described by the local equilibrium assumption eq. \eqref{eq:local_assumption} and a constitutive relation in the form of eq. \eqref{eq:constitutive_relation}, a Lyapunov function is provided by the dot product $G$ in eq. \eqref{eq:G_product}, whose non-total differential is,
\begin{equation}
	\begin{aligned}
		\frac{\mathrm{d}_X G^a}{\mathrm{d}t} &\equiv \int_V \boldsymbol{J}^a \cdot \nabla \frac{\mathrm{d}\Gamma^a}{\mathrm{d}t} \mathrm{d}V\\
		&=-\int_V \frac{\mathrm{d}\Gamma^a}{\mathrm{d}t}\nabla\cdot\boldsymbol{J}^a \mathrm{d}V\\
		&=\int_V \rho \frac{\mathrm{d}\Gamma^a}{\mathrm{d}a}\left( \frac{\partial a}{\partial t}\right)^2\mathrm{d}V \leq 0.
	\end{aligned}
	\label{eq:dxG}
\end{equation}
The validation of the inequality \eqref{eq:dxG} does not require the constitutive relations to be entropic. As an example, for the heat conduction process under the Fourier's law which is not entropic, the non-total differential of entransy dissipation rate stays negative before the stationary state is reached, i.e., $\mathrm{d}_X G^e \leq 0$. Therefore, it is inappropriate to raise such defined quantity to describe properties of entropy production as the feature is commonly shared by non-equilibrium transport processes, whose constitutive relations are not necessarily entropic.

\subsection{Numerical experiments for the heat conduction process}
\label{subsec:counterexamples}
Given the Fourier's law in the dimensionless form
\begin{equation}
	\boldsymbol{q}^* = k^* \nabla\left(-T^*\right),
\end{equation}
the dimensionless thermal diffusion equation can be written as,
\begin{equation}
	\frac{\partial T^*}{\partial t^*} = \alpha^* {\nabla^*}^2 T^*,
	\label{eq:dimensionless_diffusion}
\end{equation}
where the dimensionless conductivity $k^*=1$ and diffusivity $\alpha^*=1$ if characteristic parameters $t_0, x_0, T_0$ are properly chosen accordingly. The dimensionless EPR and entransy dissipation rate (EDR) can be respectively expressed as,
\begin{equation}
	\begin{aligned}
		P_u^* &= \int_{V^*} k^* \frac{\nabla^* T^*\cdot\nabla^* T^*}{{T^*}^2}\mathrm{d}V^*,\\
		G^{e*} &= \int_{V^*} k^* \nabla^* T^*\cdot \nabla^* T^*\mathrm{d}V^*.
	\end{aligned}
\end{equation}

To raise counterexamples that violate the principle of MinEP, initial temperature distribution $T^*(\boldsymbol{x}^*,t^*=0)$ should be given to satisfy
\begin{equation}
	\forall \boldsymbol{x}^*\in V^*, \frac{{\nabla^*}^2 T^*}{T^*}\left(\frac{{\nabla^*}^2 T^*}{T^*}-\frac{\nabla^* T^*\cdot\nabla^* T^*}{{T^*}^2}\mathrm{d}V^*\right)<0,
\end{equation}
in which case the dimensionless EPR increases according to eq. \eqref{eq:time_derivative_epr_kconst_final}. We now study the one-dimensional heat conduction process along $x^*$ axis within a unit cube. Such condition can be easily met if we take a parabola $T^*(\boldsymbol{x}^*,t^*=0)=a{x^*}^2+bx^*+c$ as the initial condition where parameters $a,b,c$ should be properly chosen. Fixing the boundary conditions as,
\begin{equation}
	\forall t^*>0,
	\left\{\begin{aligned}
		T^*(x^*=0,t^*) &= T(x^*=0,t^*=0) = c\\
		T^*(x^*=1,t^*) &= T^*(x^*=1,t^*=0)=a+b+c
	\end{aligned}\right.
\end{equation}

A set of parameters can be given as $(a,b,c)=(5,3,1)$. We calculate the evolution of dimensionless temperature field along $x^*$ axis following diffusion eq. \eqref{eq:dimensionless_diffusion}, and the dimensionless EPR and EDR are calculated accordingly.

\begin{figure}[htbp]
	\centering
	\caption{The one-dimensional heat conduction process under the Fourier's law within a unit cube, given the initial dimensionless condition $T^*(x^*,t^*=0) = a{x^*}^2+bx^*+c$ where $(a,b,c)=(5,3,1)$.}
	\label{fig:counterexample}
	\includegraphics[width=0.3\textwidth]{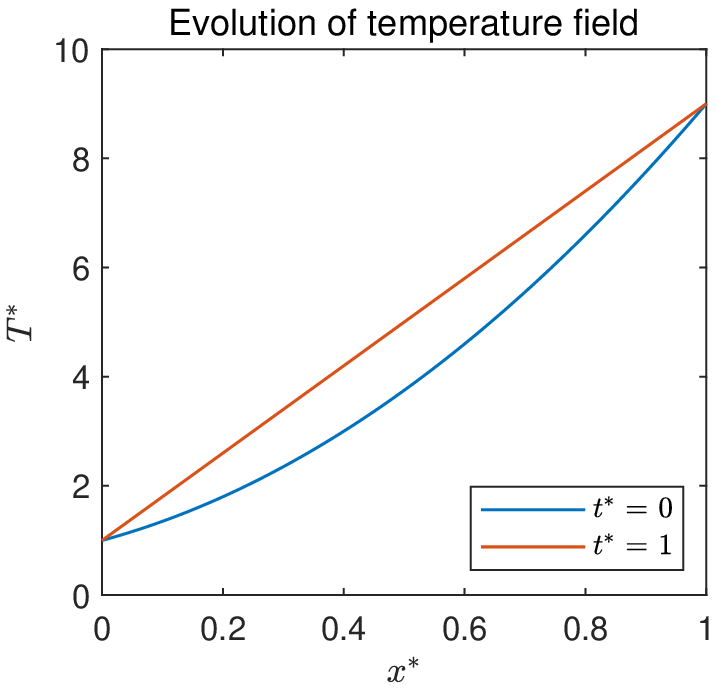}
	\includegraphics[width=0.3\textwidth]{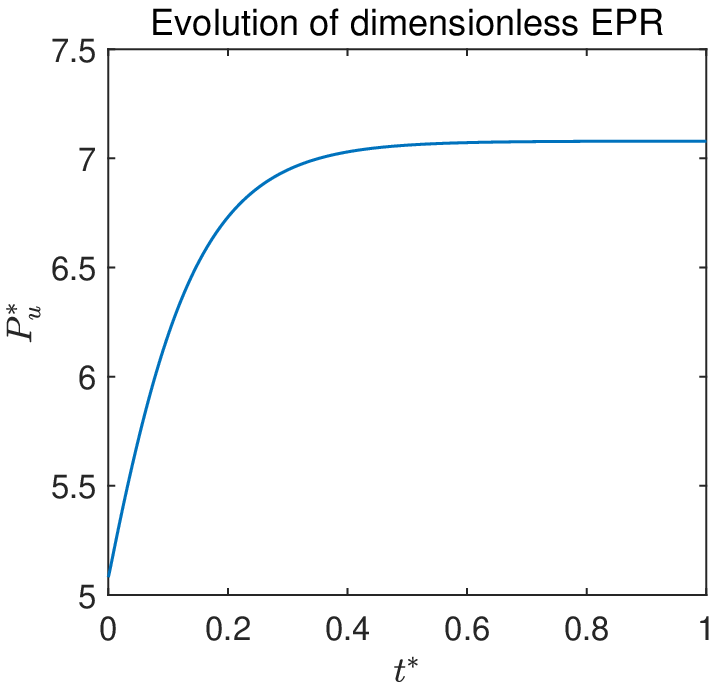}
	\includegraphics[width=0.3\textwidth]{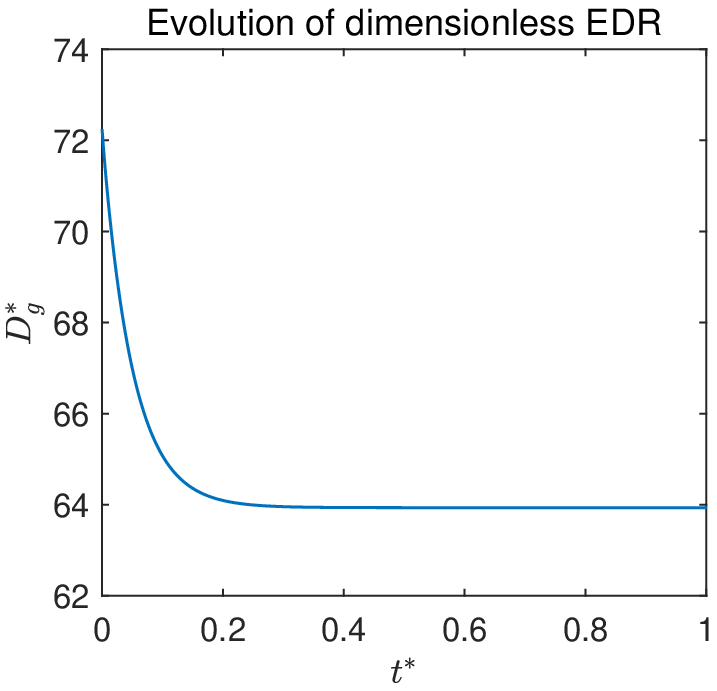}
\end{figure}

As shown in fig. \ref{fig:counterexample}, as the temperature field evolves, the EPR increases until the stationary state is reached, which violates the principle MinEP as discussed. Similar results are reported by Zullo \cite{zullo2016}, that the EPR may increase or decrease, and cannot be characterized by an extremum principle. Moreover, our numerical calculations show that the EDR decreases and is a Lyapunov function for such process, suggested by the previous theorem. 

\section{Conclusion}

\begin{enumerate}
	\item Entropy production has been long regarded as the Lyapunov functions for all kinds of non-equilibrium transport processes as the principle of MinEP claims to characterize the stationary states of irreversible processes. However, the principle of MinEP is questionable. A strong assumption is required that constitutive relations should be entropic, which is hardly corresponding to practical situations. On the other hand, such assumption of the principle is equivalent to the validation of the principle itself, which means that a circular logic is misused, and no additional information can be obtained by the principle.
	\item Attempts to modify the principle of MinEP are analyzed. Prigogine's approximation that the phenomenological coefficients of the entropic relations should be constant is inappropriate. The non-total differential, or the kinetic potential, is also inadequate as its property is commonly shard by the dot products of fluxes and forces of non-equilibrium processes, whose constitutive relations are not necessarily entropic.
	\item A general model of parabolic-like transport processes is analyzed, and a theorem is derived, demonstrating that the dot products of fluxes and corresponding forces serve as Lyapunov functions. Such fluxes and forces are provided by their actual constitutive relations (e.g., the Fourier's law, the Fick's law, the Ohm's law). Lyapunov functions for the heat conduction process, the mass diffusion process and the electrical conduction process are studied to verify the validation of the theorem. It is worth mentioning that for heat conduction (resp. electrical conduction), the entransy dissipation rate (resp. the Joule heating power), rather than the entropy production rate, serves as a Lyapunov function when the Fourier's law (resp. the Ohm's law) is valid.
	\item Numerical experiments for the heat conduction process are effectuated to verify the validation of the theorem. It demonstrates that the principle of MinEP will be violated when the Fourier's law is valid.
\end{enumerate}


\end{document}